# A Low-cost Quasi-planar Array Probe for Photoacoustic Imaging

Xiyu Chen,[1,†] Junxiang Cai,[1,†] Rui Zheng,[1] Tao Wu,[1,*] And Fei Gao,[2,3,4,*]

[1]*School of Information Science and Technology, ShanghaiTech University. Shanghai, China*
[2]*School of Biomedical Engineering, Division of Life Sciences and Medicine, University of Science and Technology of China, Hefei, Anhui, 230026, China*
[3]*Hybrid Imaging System Laboratory, Suzhou Institute for Advanced Research, University of Science and Technology of China, Suzhou, Jiangsu, 215123, China*
[4]*School of Engineering Science, University of Science and Technology of China, Hefei, Anhui, 230026, China*

[†]*These authors contributed equally to this work.*
*\*wutao@shanghaitech.edu.cn; fgao@ustc.edu.cn*

**Photoacoustic imaging (PAI) is a novel hybrid imaging technique that combines the benefits of both optical and acoustic imaging modalities, which provides functional and molecular optical contrasts of deep tissue. Commonly used ultrasound transducers for PAI include linear and planar arrays, which can provide two-dimensional (2D) and three-dimensional (3D) image reconstruction, respectively. However, linear arrays cannot provide reconstruction of 3D images, which makes it impossible to locate chromophores in 3D space. Although planar array can provide fast 3D imaging in real time, it usually requires thousands of analog-to-digital conversion channels for data acquisition, which is costly. To fill the gap between 2D and 3D PAI, we propose a quasi-planar array that uses double 16-elements-linear arrays arranged in parallel to achieve real-time 3D imaging. We first conducted simulation studies to prove that the quasi-planar probe can perform 3D imaging to localize simple chromophores. Then, the agarose phantom experiment demonstrated that the probe can reconstruct 3D imaging of multiple absorbers in different depths. A potential application of this device is to provide a low-cost 3D PAI solution for fast tracking of needle tip during needle biopsy, which will be further explored in our future work.**

Photoacoustic imaging (PAI) is a novel imaging method that harnesses the strengths of both optical and acoustic imaging modalities, offering functional and molecular optical contrasts for deep tissue analysis [1-3]. One of the key advantages of photoacoustic imaging is its ability to preserve high acoustic resolution in deep tissue, overcoming the limitations imposed by light scattering. In a wide range of clinical applications, accurate placement of medical devices such as needles and catheters is critical to obtain better patient outcomes. In recent years, three-dimensional (3D) photoacoustic imaging (PAI) based on two-dimensional matrix arrays has rapidly developed and shown great potential for in-vivo imaging. Researchers have developed various array configurations, such as spherical [4-6] and planar arrays [7-9], and also have used a non-focused planar array to achieve 3D photoacoustic imaging by translating and synthesizing planar arrays [10]. However, 3D PAI requires a very dense acoustic sensor configuration, typically several thousands of elements, to achieve high-quality imaging, which results in extremely high system costs and complexity, or significant time costs associated with the synthetic array that needs to be mechanically scanned.

Interventional ultrasound (US) is increasingly and widely applied in modern clinical medicine. Nevertheless, interventional US often fails due to either the specular reflection from the needle's smooth surface or the excessive mismatch of acoustic impedance between the needle and surrounding heterogeneous tissue. PAI shows promise for alternatively visualizing puncture needle [11, 12]. The constituent materials of most commercial needle, i.e. metal, can absorb the energy from the pulse laser illumination, which then generate ultrasonic signal based on the thermoelastic effect [13]. In order to achieve the goal of navigation, we focus our attention on positioning the tip of the endoscopic probe. Considering the point-like morphology of the target object, we need to achieve needle tip's localization in 3D space in real time, without the requirements for detailed 3D imaging of the whole volume. To address this clinical need, we demonstrated a quasi-planar probe, which can potentially provide positioning navigation for interventional needles at a much lower system cost.

In the former work, we made a 16-channels linear array PA probe [14], shown in Figure 1(a), where the laser fiber connector and electrical cable connector are used to input the laser illumination and output the PA signal. The light output clings to the transducer array and emits laser light perpendicular to the ultrasonic transducer array's receiving surface. As shown in Fig. 1(b), the acoustic receiving angle of the ultrasonic transducer is 14 degrees, forming a fan-shape imaging plane parallel to the laser output plane. In this work, we integrated two 16-channels linear arrays to build a quasi-planar array, forming a 2×16 quasi planar array. Figure 1(c) shows the photograph of the proposed PA probe. Figure 1(d) shows the detail layout of the detection end of the probe. Two sets of 16-channels linear transducers are placed side by side, optical fibers are led out from the gap between the two sets of transducers for light illumination. Now, due to the parallel arrangement of two linear arrays, we can treat such configuration as quasi-planar, making 3D image reconstruction possible. Meanwhile, due to the use of only two parallel linear array transducers with only 16 elements, we also ensure the extremely low-cost advantage of the proposed probe.

Fig. 1. Photographs of two versions of the probe. (a) Photograph of the PAI probe system; (b) Schematic of the distal end of the probe. (c) Photograph of the improved probe in this work; (d) Details of the quasi-planar array.

We use only two sets of linear arrays in the Y axis to build the quasi-planar array, ensuring the minimal implementation cost of 3D imaging reconstruction. We fabricated the device by integrating a custom-made plastic optical fiber and a designed printed circuit board that contains signal amplifier circuits, and bonding two 16-elements linear transducer arrays.

We conducted simulation study using k-wave [15]. The position of the transducer elements was set according to the actual size parameters of the probe, and a dot-like sound source was set to mimic the photoacoustic point source. By changing the position of the voxel of the sound source, we obtained 3D reconstruction renderings tracking this point-like source, which was mimicking the needle tip. Figure 2(a) shows five different position settings of dot sound source. Then, delay and sum algorithm was used to reconstruct PAI images in 3D. We filtered and weighted the reconstruction results, retaining voxels with signal strength greater than 50% of the maximum value and filtering out the 50% of low signal voxels, thereby reducing the impact of inherent artifacts induced by the DAS algorithm due to sparse view and limited angle issues. Figure 2(b) shows the reconstructed 3D PA images. It can be seen that there are obvious bright spots in the 3D images. The positions of the bright spots in these images match the sound source position well. The simulation experiments proved that the designed quasi-planar array can perform spatial positioning on simple chromophores. It can be seen that the quasi-planar array PAI system can effectively track and locate the spatial position of the sound source.

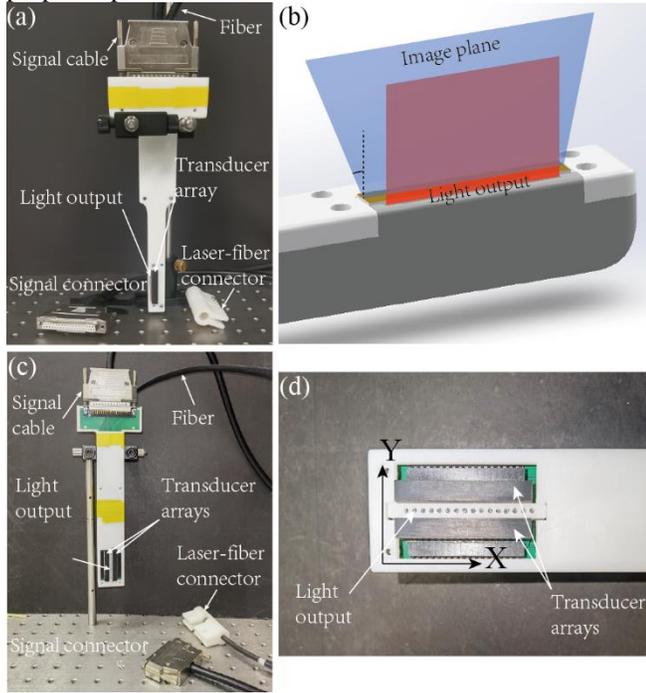

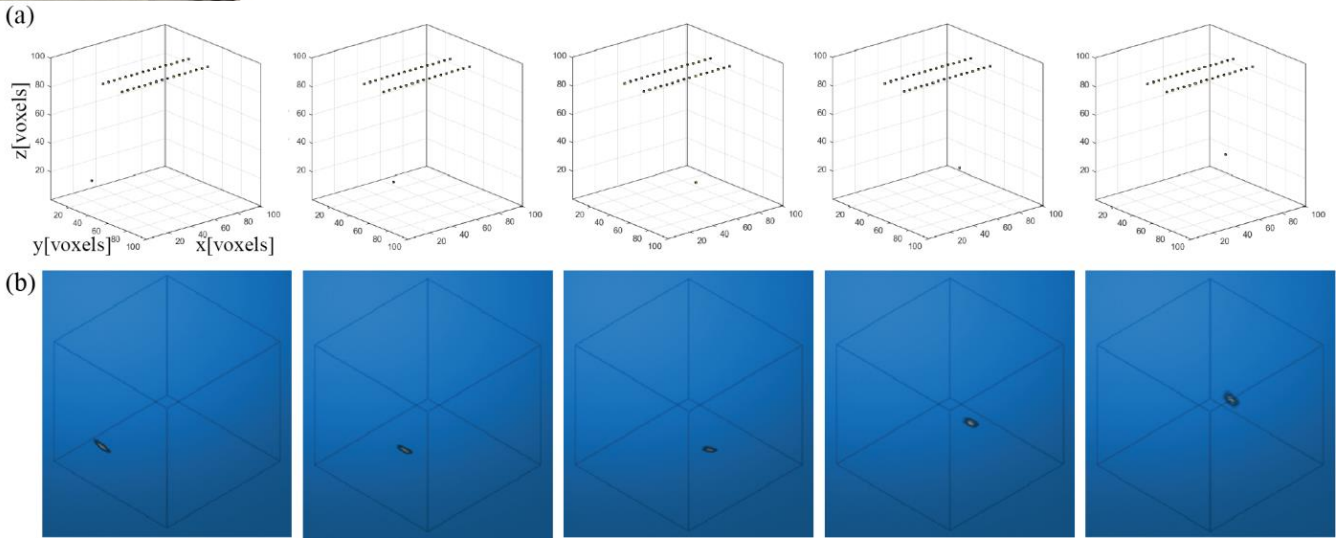

Fig. 2. Simulation experiments of 3D tracking using the proposed quasi-planar PA probe. (a) Five different position settings of dot-like sound source. (b) Simulation experiments 3D reconstructed images by DAS.

We built a photoacoustic imaging system based on our proposed quasi-planar probe to acquire the positioning and navigation images of the intervention needle. The system contains a pulse laser source (PhotoSonus Series Tunable Nd:YAG Laser System, Ekspla), a data acquisition device (Flash DAQ32, PhotoSound), the quasi-planar array probe and a computer.

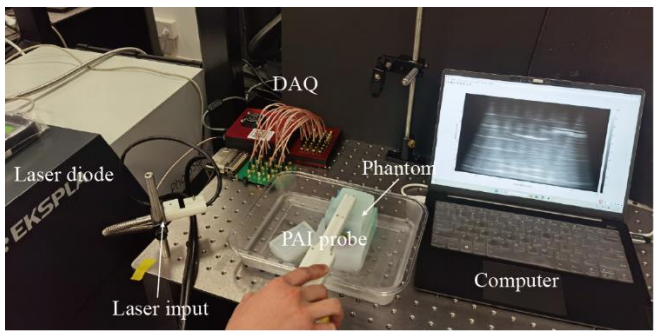

Fig. 3. The photograph of the quasi-planar array PAI system setup.

The laser source generates pulsed laser with 20 Hz repetition rate and 10 ns pulse width. The laser source also provides synchronous trigger signal to synchronize data acquisition device. The quasi-planar array probe detected the 32 channels' photoacoustic signal, which are fed into the data acquisition device followed by computer. The computer is synchronized with the laser firing through communication with data acquisition device for the PA data transmission and storage. Image reconstruction using 3D delay and sum (3D-DAS) algorithm is performed in the computer. The system setup is shown in Figure 3.

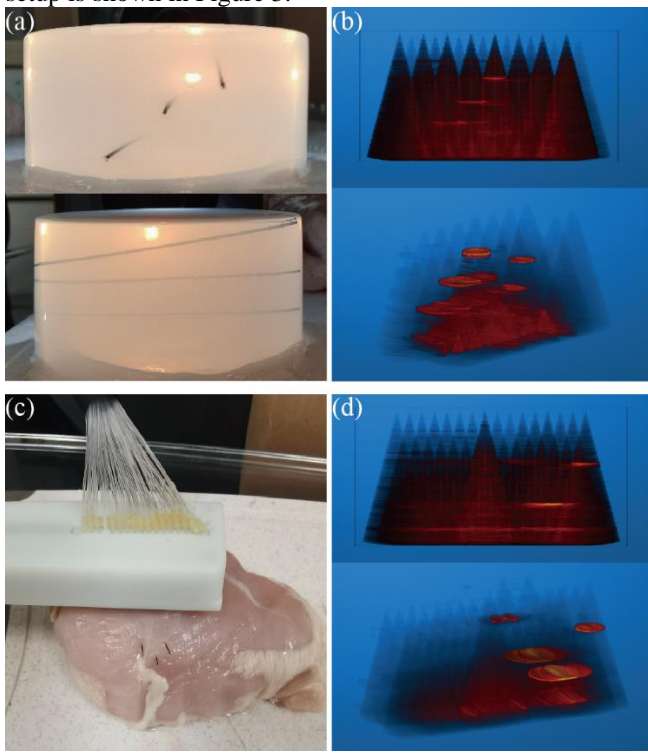

Fig. 4. (a) Pencil lead phantom (0.5mm) in agar, at three different depths and in different longitudinal planes. (b) Pencil lead – agar phantom images reconstructed by 3D DAS algorithm, front and oblique views. (c) Muscle tissue phantom containing foreign object to be tested. Chicken breast meat containing three pieces of black iron wires (0.5mm) placed at different depths. (d) Chicken breast meat – black irons wires phantom images reconstructed by 3D DAS algorithm, front and oblique views.

In order to evaluate the deep-tissue imaging capabilities of the quasi-planar array PAI system, several experimental samples were prepared and imaged. To demonstrate the high geometric uniformity of the system, a pencil leads phantom embedded in agar was imaged (Figure. 3(a)). Figure. 3(b) shows the reconstructed images using 3D-DAS algorithm. Images reconstructed from 3D-DAS, including front and oblique views. The front view clearly shows three layers of bright spots. By numerical processing and filtering out background noise, six distinct bright spots can be seen, with each pair representing the position of a pencil lead. We can see the distinct bright spots, which displayed three pencil leads at different depths that corresponded well with the sample, proving the system's ability to accurately capture tissue structures at different depths. Due to the scarcity of channels, there are only 16 and 2 channels on the x and y axes respectively, which inevitably makes these light spots appear relatively large. Due to the signal coefficient, the delay and sum algorithm inevitably produces large arc-shaped bright spots. However, we can clearly distinguish different layer thicknesses and the location of bright spots within the plane, which can assist us in determining the position of the pencil lead.

To simulate the real situation of the probe swimming in muscle tissue, three black iron wires with a diameter of 0.5mm in the chicken breast meat were placed in the chicken breast meat (Figure. 3(c)). Figure. 3(d) shows the reconstruction effect in the replica of the actual situation. In the front view, we can see two bright spots and one relatively blurry spot, which is due to the different positioning of the wire. The photoacoustic signal source generated by the wire at the rear is not as clear in the front view, but from the oblique view, we can clearly see three bright spots at different positions, representing the different positions of the three wires. Due to being a phantom made of chicken breast, the energy attenuation of the laser beam in muscle tissue transmission is faster, inevitably leading to a decrease in signal-to-noise ratio. Therefore, we can also see some other noise generated artifacts, but relatively brighter positions can still be well distinguished.

These experimental results highlight the potential of the quasi-planar array probe system in achieving high-resolution imaging and positioning, navigation function of the interventional needle during surgery. In conclusion, we improved the handheld linear array probe with low cost transducer to the quasi-planar array probe. The probe contains 32 channels, consist two sets of 16 channels linear array. Through numerical simulation and simulated experiments, the handheld low channel quasi-planar array probe system achieves 3D detection and reconstruction ability, we have demonstrated the feasibility and accuracy of our system. Although the quality of 3D reconstruction is limited by the sparse data volume caused by fewer channels and the lower resolution resulting from DAS algorithms, it is sufficient to meet the expected spatial positioning capability. Moreover, fewer channels can effectively control costs, laying the foundation for practical applications. As the consequence, the proposed low-cost quasi-planar array system shows promising prospect for the clinical applications, e.g.

navigation of intraoperative needle intervention etc., which will be further validated in vivo in the future work.

**Acknowledgment.**

**References**

1. L. V. Wang and S. Hu, "Photoacoustic Tomography: In Vivo Imaging from Organelles to Organs," Science **335**, 1458-1462 (2012).
2. A. Dangi, et al., "Towards a low-cost and portable photoacoustic microscope for point-of-care and wearable applications," IEEE Sensors Journal **20**, 6881-6888 (2019).
3. Y. Liang, et al., "Optical-resolution functional gastrointestinal photoacoustic endoscopy based on optical heterodyne detection of ultrasound," Nature Communications **13**, 7604 (2022).
4. L. Lin, et al., "High-speed three-dimensional photoacoustic computed tomography for preclinical research and clinical translation," Nature communications **12**, 882 (2021).
5. Y. Matsumoto, et al., "Label-free photoacoustic imaging of human palmar vessels: a structural morphological analysis," Scientific reports **8**, 786 (2018).
6. I. Ivankovic, et al., "Real-time volumetric assessment of the human carotid artery: handheld multispectral optoacoustic tomography," Radiology **291**, 45-50 (2019).
7. W. Kim, et al., "Wide-field three-dimensional photoacoustic/ultrasound scanner using a two-dimensional matrix transducer array," Optics Letters **48**, 343-346 (2023).
8. D. Piras, et al., "Photoacoustic imaging of the breast using the twente photoacoustic mammoscope: present status and future perspectives," IEEE Journal of Selected Topics in Quantum Electronics **16**, 730-739 (2009).
9. M. Heijblom, et al., "Visualizing breast cancer using the Twente photoacoustic mammoscope: what do we learn from twelve new patient measurements?," Optics express **20**, 11582-11597 (2012).
10. S. Li, et al., "Photoacoustic imaging of peripheral vessels in extremities by large-scale synthetic matrix array," Journal of Biomedical Optics **29**, S11519-S11519 (2024).
11. T. Zhao, et al., "Minimally invasive photoacoustic imaging: Current status and future perspectives," Photoacoustics **16**, 100146 (2019).
12. J. Su, et al., "Photoacoustic imaging of clinical metal needles in tissue," Journal of biomedical optics **15**, 021309-021309-021306 (2010).
13. D. Piras, et al., "Photoacoustic needle: minimally invasive guidance to biopsy," Journal of biomedical optics **18**, 070502-070502 (2013).
14. J. Cai, et al., "A Low-cost High-sensitivity Endoscopic Probe for Real-time Photoacoustic Imaging," in *2023 IEEE International Ultrasonics Symposium (IUS)*, (IEEE, 2023), 1-3.
15. B. E. Treeby and B. T. Cox, "k-Wave: MATLAB toolbox for the simulation and reconstruction of photoacoustic wave fields," Journal of biomedical optics **15**, 021314-021314-021312 (2010).